\documentclass[11pt,dvips]{article}
\usepackage{epsfig,times} 
\usepackage{picinpar}
\usepackage{wrapfig}
\usepackage{floatflt}
\makeatletter
\renewcommand{\section}{\@startsection{section}{1}{0in}
	{0.4\baselineskip}{0.1\baselineskip}{\Large\bf}}
\renewcommand{\subsection}{\@startsection{subsection}{2}{0in}
	{0.25\baselineskip}{-\baselineskip}{\large\bf}}
\renewcommand{\subsubsection}{\@startsection{subsubsection}{3}{0in}
	{0.1\baselineskip}{-\baselineskip}{\normalsize\bf}}
\makeatother
\begin{document}

\thispagestyle{myheadings}
\markright{HE 1.1.14}
\begin{center}
{\LARGE \bf  Leading nucleon and the proton-nucleus inelasticity}
\end{center}

\begin{center}
{\bf J. Bellandi$^{1}$, J.R. Fleitas$^{1}$, and J. Dias de Deus$^{2}$}\\
{\it $^{1}$Instituto de F\'{\i}sica, Universidade Estadual de Campinas, Campinas, SP 13083-970, Brazil\\
$^{2}$Instituto Superior T\'ecnico - CENTRA, Av. Rovisco Paes, 1, 1096 Lisboa Codex, Portugal}
\end{center}

\begin{center}
{\large \bf Abstract\\}
\end{center}
\vspace{-0.5ex}

We present in this paper, a calculation of average proton-nucleus inelasticity. Using an iteractive leading particle model and the Glauber model,  we relate the leading particle distribution in nucleon-nucleus iteractions with the respective one in nucleon-proton collisions. To describe the leading particle distribution in nucleon-proton, we use the Regge-Mueller formalism.

\vspace{1ex}

We calculate the average proton-nucleus inelasticity. Using an Iterative Leading Particle Model (Frichter et al., 1997) and the Glauber
model  (Glauber, 1959; Glauber et al., 1970), we relate the leading particle distribution in nucleon-nucleus
interactions with the respective one in nucleon-proton collisions. In this
model the leading particle spectrum in {\it \ $p+A\rightarrow N$}(nucleon)$+$%
{\it $X$} collisions is built from sucessive interacions with $\nu $
interacting proton of the nucleus $A$ and the behaviour is controlled by a
straightforward convolution equation. It should be mentioned that, strictlly
speaking, the convolution should be 3-dimensional. Here we only considered
the 1-dimension approximation. In a recent paper (Bellandi et al., 1999) we have used this model
to describe the hadronic flux in the atmosphere, showing that the average
nucleon-nucleus elasticity, $<x>_{N-A}$, is correlated whit the respective
average nucleon-proton elasticity, $<x^\gamma >_{N-p}$, by means of the
following relation

\begin{equation}
\label{1}(1-<x>_{N-A})=\frac 1{\sigma _{in}^{N-ar}}\int d^2b\left[ 1-\exp
[-(1-<x>_{N-p})\sigma _{tot}^{pp}T(b)]\right] 
\end{equation}
where $T(b)$ is the nuclear thickness and given by means of the Woods-Saxon
model (Woods, \& Saxon, 1954; Barrett, \& Jackson, 1977). Introducting the inelasticity given by $<k>=1-<x>,$ this
expression can be transformed in 
\begin{equation}
\label{2}<k>_{N-A}=\frac 1{\sigma _{in}^{N-A}}\int d^2b\left[ 1-\exp
[-<k>_{N-p}\sigma _{tot}^{pp}T(b)]\right] .
\end{equation}

It is clear from this relationship that only in small $\sigma _{tot}^{pp}$
limit is $<K>_{N-A}\simeq <K>_N$ . In general, $<K>_{N-A}\geq <K>_N$ and the
effect increasing with the increase of $\sigma _{tot}^{pp}.$ If $%
<K>_N\rightarrow 0,$ one also has $<K>_{N-A}\rightarrow 0.$ On the other
hand, if $<K>_N=1$, then $<K>_{N-A}=1,$ and Eq. (\ref{2}) reproduces the
Glauber model relationship between $\sigma _{in}^{N-A}$ and $\sigma
_{tot}^{pp}.$ In order to calculate $<K>_{N-A}$ we use for $<K>_N$ the
values calculated by means of the Regge-Mueller formalism (Batista et al., 1998) and as input
for $\sigma _{tot}^{pp}$ we have used the UA4/2 parametrizations for the
energy dependence (Burnett et al., 1992). In the Fig. (1) we show the results of this calculations
for the following nuclei: C, Al, Cu, Ag, Pb and air ($A=14.5$). In this figure we
also show recent experimental data for p-Pb, $<K>=0.84\pm 016$ (Barroso et al., 1997) and for
p-C, $<K>=0.65\pm 0.08$ (Wilk \& Wlodarczyk, 1999).

In the Fig. (2), we compare the calculated $<K>_{p-air}$ with results from some
models used in Monte Carlo simulation (Gaisser et al., 1993); the Kopeliovich {\it et al.} (Kopeliovich et al., 1989) (KNP)
QCD multiple Pomeron exchanges model; the Dual Parton model with sea-quark
interaction of Capella {\it et al.} (Capella et al., 1981); the statistical model of Fowler {\it et al.}
(Fowler et al., 1987) and with calculated values derived from cosmic ray data by Bellandi {\it et
al.} (Bellandi et al., 1998). We note that the calculated $<K>_{p-air}$  (Bellandi et al., 1998) was done
assuming for the $T(b)$ nuclear thickness the Durand and Pi model (Durand \& Pi, 1988), which
gives small values for the average inelasticity. In the Fig. (2) we also show
the average inelasticity values as calculated by means of this model.
\begin{figure}
\begin{center}
{\mbox{\epsfig{file=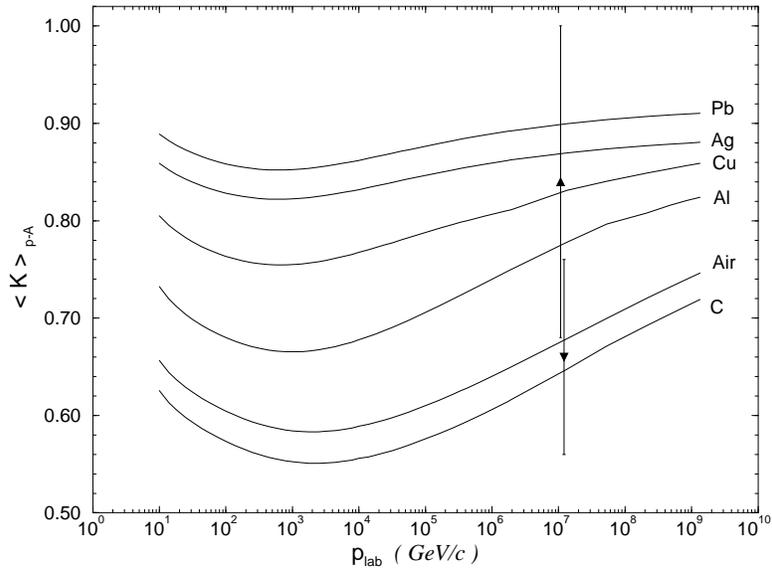,width=4.8in}}}
\end{center}
\vspace{-4ex}
\caption{ Proton-nucleus inelasticities calculated. The up triangle is Pb data (Barroso et al., 1997) and down triangle is C data (Wilk \& Wlodarczyk, 1999) .}
\end{figure}

\vspace{0.5cm}

\begin{figure}
\begin{center}
{\mbox{\epsfig{file=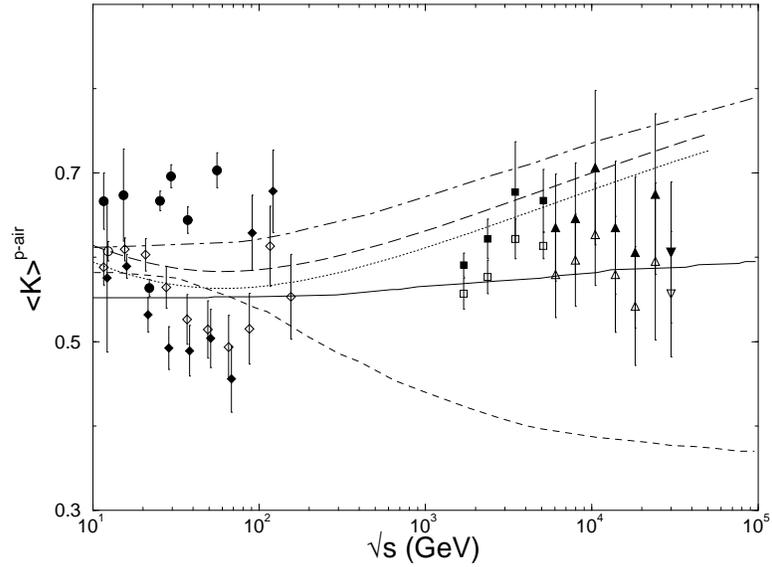,width=4.8in}}}
\end{center}
\vspace{-4ex}
\caption{ The $ <K>^{p-air} $ as a function of $ \sqrt{s} $ in $ GeV $. The experimental data from (Bellandi et al., 1998). Dash line from (Fowler et al., 1985). Solid line from (Capella et al., 1981). Dot-dash line from (Kopeliovich et al., 1989). Dot line from Eq. (2) with Woods-Saxon model  (Woods \& Saxon, 1954; Barrett, \& Jackson, 1977). Long-dash line from Durand-Pi model (Durand \& Pi, 1988).}
\end{figure}

\vspace{1ex}

We would like to thank the Brazilian governmental agencies CNPq and CAPES
for financial support

\medskip 

%
\vspace{1ex}
\begin{center}
{\Large\bf References}
\end{center}
Barrett, R.C. \& Jackson, D.F. 1977,  Nuclear Sizes and Structure, (Clarendon Press, Oxford). \\
Barroso, S.L.C. {\it et al}. (Chacaltaya Collab.) 1997,  Proc. 25$^{\rm th}$ ICRC, (Durban, 1997) \\
Batista, M. \& Covolan, R.J.M. 1998, hep-ph:9811425. \\
Bellandi, J., Fleitas, J.R. \& Dias de Deus, J. 1998, Il Nuovo Cimento, {\bf 111A}, 149. \\
Bellandi, J. et al. 1999, Proc. 26$^{\rm th}$ ICRC, (Salt Lake City, 1999). \\
Burnett, T.H. {\it et al.}, (UA4/2 Collab.) 1992, ApJ {\bf 349} L25. \\
Capella, A. {\it et al.} 1981, Z. Phys. {\bf C 10}, 249. \\
Durand, L. \& and Pi, H. 1988, Phys. Rev. {\bf D 38}, 78. \\
Fowler, G. {\it et al.} 1987, Phys. Rev. {\bf D 35}, 870. \\
Frichter, G.M. {\it et al.} 1997, Phys. Rev. {\bf D 56}, 3135. \\
Gaisser, T.K. {\it et al.} 1993, Phys. Rev. {\bf D 47}, 1919. \\
Glauber, R.J. 1959, Lect. Theor.  Phys. Vol.1, edited by W.Britten and L.G.Dunhan (Interscience, NY), 135. \\
Glauber, R.J. {\it et al.} 1970, Nucl. Phys.{\bf B 12}, 135.\\
Kopeliovich, B.Z. {\it et al.} 1989, Phys. Rev. {\bf D 39}, 769. \\
Wilk, G. \& Wlodarczyk, Z. 1999, Phys. Rev. {\bf D 57}, 180. \\
Woods, R.D. \& Saxon, D.S. 1954, Phys. Rev. {\bf 95}, 577.

{\it Contribution to the 26$^{\rm th}$ International Cosmic Ray Conference, Salt Lake City, Utah - August, 1999}

\end{document}